\documentclass[
twocolumn,
superscriptaddress,
amsmath,amssymb,
aps
]{revtex4-2}

\usepackage{graphicx}
\usepackage{bm}
\usepackage{hyperref}
\usepackage{physics}
\usepackage{mathtools}
\usepackage{mathrsfs}
\usepackage{siunitx}
\usepackage{textcomp}
\usepackage{orcidlink}

\DeclareFontFamily{U}{mathx}{\hyphenchar\font45}
\DeclareFontShape{U}{mathx}{m}{n}{<-> mathx10}{}
\DeclareSymbolFont{mathx}{U}{mathx}{m}{n}
\DeclareMathAccent{\widebar}{0}{mathx}{"73}

\newcommand{\schro}{Schr{\"o}dinger~}
\usepackage[normalem]{ulem}
\usepackage{xcolor}
\definecolor{amber}{rgb}{1,0.49,0}
\definecolor{darkgreen}{rgb}{0,0.55,0}
\definecolor{tangerine}{rgb}{0.944,0.522,0}
\definecolor{verde}{rgb}{0.,0.6,0}
\definecolor{rosso}{rgb}{0.9,0.0,0.2}
\definecolor{magenta}{rgb}{0.9,0.2,0.9}

\newif\ifhighlight

\newcommand{\highlight}{\highlighttrue}
\highlight
\newcommand{\editor}[2]{%
  \expandafter\newcommand\csname #1note\endcsname[1]{%
    \textcolor{#2}{(\textbf{#1:} ##1)}}%
  \expandafter\newcommand\csname #1\endcsname[1]{%
    \ifhighlight\textcolor{#2}{##1} \else ##1\fi}%
  \expandafter\newcommand\csname #1cancel\endcsname[1]{%
    \ifhighlight\textcolor{#2}{\sout{##1}}\fi}%
  \expandafter\newcommand\csname #1change\endcsname[2]{%
    \ifhighlight\textcolor{#2}{\sout{##1} ##2}\else ##2\fi}%
  \newenvironment{#1text}{\ifhighlight\color{#2}\fi}{\color{black}}
}

\editor{resub}{blue}
\editor{AF}{cyan}
\editor{NM}{purple}
\editor{CAVEAT}{red}

\begin{document}

\title{Quantum annealing for materials }

\author{Alfredo Fiorentino\,\orcidlink{0000-0002-3048-5534}}
\email{alfredo.fiorentino@psi.ch}
\affiliation{PSI Center for Scientific Computing, Theory and Data, 5232 Villigen PSI, Switzerland}
\affiliation{National Centre for Computational Design and Discovery of Novel Materials (MARVEL), 5232 Villigen PSI, Switzerland}

\author{Nicola Marzari\,\orcidlink{0000-0002-9764-0199}}
\affiliation{PSI Center for Scientific Computing, Theory and Data, 5232 Villigen PSI, Switzerland}
\affiliation{National Centre for Computational Design and Discovery of Novel Materials (MARVEL), 5232 Villigen PSI, Switzerland}
\affiliation{U Bremen Excellence Chair, Bremen Center for Computational Materials Science, and MAPEX Center for Materials and Processes, Universität Bremen, 28359 Bremen, Germany}
\affiliation{Theory and Simulation of Materials (THEOS), and National Centre for Computational Design and Discovery of Novel Materials (MARVEL), École Polytechnique Fédérale de Lausanne, 1015 Lausanne, Switzerland}
\affiliation{Theory of Condensed Matter, Cavendish Laboratory, University of Cambridge, Cambridge, CB3 0US, United Kingdom}

% If not applicable, delete \equalauthors entirely.
% \equalauthors{\textsuperscript{1}A.F. and N.M. contributed equally to this work.}

% \correspondingauthor{\textsuperscript{1}To whom correspondence should be addressed. E-mail: alfredo.fiorentino@psi.ch}


\begin{abstract}
Finding the global minimum of a potential energy surface is a fundamental challenge in materials science, with applications ranging from protein folding to cluster physics and, more broadly, to systems in which the number of (meta)stable configurations grows prohibitively large. In recent decades, quantum annealing (QA) has emerged as a promising global optimization strategy, exploiting quantum fluctuations in contrast to the thermal fluctuations that drive its classical counterpart.

Here, we introduce a novel implementation of QA based on path-integral molecular dynamics, an efficient and well-established framework for sampling the quantum nuclear density without the need to manipulate many-body wavefunctions explicitly. While retaining the flexibility and simplicity of molecular dynamics simulations, this quantum-annealing protocol delivers strong performance across a wide range of atomic systems, simulated by either empirical force fields or machine-learning interatomic potentials. The method can be used either as a global optimizer of the potential-energy surface, or as a quantum-informed structure-search strategy in which nuclear quantum effects are included directly in the optimization workflow---a feature particularly relevant for materials such as high-pressure hydrides.
\end{abstract}

\maketitle

\section{Introduction}
% \AF{Comandi per aggiunte,} \AFnote{note}, \AFchange{cambi}{sostituzioni}, \AFcancel{tagli}%%% AF per I comandi di Alfredo, NM per Nicola.
The search for the global minimum (GM) of a complex potential energy surface (PES) is a central problem in physics, chemistry, and materials science. Many relevant systems—from biomolecules and glasses to atomic and molecular clusters—exhibit rugged energy landscapes characterized by an exponentially large number of metastable states. While the choice of an optimal optimization strategy generally depends on the specific problem, simulated annealing (SA) remains a time-honored and widely used approach \cite{press1992numerical,kirkpatrick1983optimization}, valued for its generality and flexibility. Inspired by metallurgical practices, SA mimics a melting phase followed by slow cooling, using thermal fluctuations to explore configurational space. However, thermal fluctuations are not the only possible driving force, and it is well known that SA can struggle in the presence of large energy barriers and multi-funnel PESs. In an effort to overcome these limitations, an alternative paradigm has been introduced: quantum annealing (QA) \cite{kadowaki1998quantum,brooke1999quantum,farhi2000quantum,amara1993global}. Drawing inspiration from quantum mechanics, QA relies on the idea that quantum fluctuations enable the exploration of classically rare or forbidden configurations, and can therefore outperform SA in certain scenarios \cite{santoro2002theory,stella2005optimization,santoro2006optimization,stella2006monte}.

QA has become a popular tool for optimizing discrete problems, such as disordered Ising models, with implementations on both classical computers \cite{amara1993global,kadowaki1998quantum,santoro2002theory,stella2005optimization,lami2023quantum} and quantum hardware—where it is often referred to as adiabatic quantum computing \cite{farhi2000quantum,johnson2011quantum,perdomo2012finding,barends2016digitized}. By contrast, its application to multidimensional continuous systems, which is the focus of this work, has remained comparatively limited despite early successes \cite{amara1993global,finnila1994quantum,gregor2005minimization}. 
A major obstacle lies in the need to represent and evolve a many-body wavefunction according to the time-dependent Schrödinger equation, either in real or imaginary time. Several strategies have been proposed to address this challenge.

These approaches can be broadly divided into two categories. The first consists of assuming a specific ansatz for the wavefunction and evolving it according to the \schro equation~\cite{amara1993global}. Alternatively, one may rely on stochastic techniques such as diffusion Monte Carlo (DMC)~\cite{finnila1994quantum} or path-integral Monte Carlo (PIMC)~\cite{gregor2005minimization} to sample the quantum nuclear density and thereby capture equilibrium nuclear quantum effects (NQEs), at the expense of abandoning exact quantum dynamics. 

Both routes require considerable care. In the former case, the ansatz must remain sufficiently flexible throughout the adiabatic evolution. In the latter, the choice of Monte Carlo moves and the associated fictitious dynamics can strongly influence—and even qualitatively alter—the performance of the algorithm~\cite{stella2006monte,inack2015simulated}.

In this work, we propose a novel implementation of QA based on path-integral molecular dynamics (PIMD). While conceptually related to earlier QA-PIMC approaches~\cite{gregor2005minimization,santoro2006optimization}, QA-PIMD preserves the simplicity and scalability of molecular dynamics simulations, while sampling the configurational space according to the path-integral approximation to the quantum nuclear density, which becomes exact in the limit of infinitely many beads and sufficient equilibration. When applied to standard optimization benchmarks such as Lennard--Jones (LJ) clusters~\cite{hoare1976statistical,northby1987structure,wales1997global}, the algorithm efficiently locates the global minimum for cluster sizes at least as large as those considered in previous QA studies~\cite{amara1993global,finnila1994quantum,gregor2005minimization}. Moreover, for these systems, QA-PIMD often shows a clear performance advantage over its classical counterpart, SA, reaching the reference minimum with shorter annealing schedules. This advantage becomes even more pronounced when the replica-pinned version of QA (RPQA)~\cite{gregor2005minimization} is used, which consistently solves the challenging $\mathrm{LJ}_{38}$ benchmark.

Beyond its role as a global optimization tool, a quantum algorithm such as QA-PIMD is particularly appealing for materials science applications. For certain materials like high-pressure hydrides~\cite{drozdov2019superconductivity,errea2016quantum,monacelli2023quantum,errea2020quantum,poletaev2025sscha,smith2026capturing}, the inclusion of NQEs—such as zero-point energy contributions—can modify the relative stability of competing structures \cite{calvo2001quantum,errea2020quantum}. In such cases, a straightforward minimization of the PES may lead to incorrect structural predictions or phase diagrams. By contrast, an optimization strategy that inherently accounts for quantum nuclear effects can naturally identify physically meaningful structures and stability rankings.

The remainder of the article is organized as follows. Section~\ref{sec:results_discussion} presents the results and discussion. We first introduce the theoretical framework underlying quantum annealing and its path-integral molecular dynamics implementation. We then assess the ability of QA-PIMD to adiabatically follow the quantum nuclear density using an asymmetric double-well potential. Next, we compare the performance of QA-PIMD, its replica-pinned variant, and (classical) simulated annealing for Lennard-Jones clusters. We also demonstrate the applicability of the method to reconstruction problems in crystalline materials described by machine-learning interatomic potentials (MLIPs), a problem of practical relevance in experimental materials characterization.
Then, the QA-PIMD is successfully applied to study structural transitions driven by NQEs, in particular for the superconducting high-pressure $\mathrm{LaH}_{10}$ structure.
Finally, section~\ref{sec:conclusions} concludes the paper with a summary and outlook. Technical and methodological details are provided in the \emph{Materials and Methods} section.

\section{Results and Discussion}\label{sec:results_discussion}

\subsection*{Quantum annealing framework}
Quantum annealing is a computational technique aimed at identifying low-energy configurations of complex systems by exploiting quantum fluctuations. The algorithm consists of adiabatically evolving the wavefunction $\psi$ according to the time-dependent Schr\"odinger equation, written here in the position representation:
\begin{align}\label{eq:bare_QA}
    &i\hbar \frac{\partial}{\partial t} \psi(\mathbf{x},t) = \hat{H}(t)\psi(\mathbf{x},t)\nonumber\\
    &\hat{H}(t) = -\alpha(t) \sum_{i=1}^{N} \frac{\hbar^2}{2m_i} \nabla_i^2 
    + U(\mathbf{x}_{1}, \ldots, \mathbf{x}_{N}),
\end{align}
where $U$ denotes the interaction potential of an $N$-body system, $\hbar$ is the reduced Planck constant, $m_i$ are the particle masses, and $\mathbf{x}$ is a shorthand notation for the full set of particle coordinates, $\mathbf{x}=\mathbf{x}_{1}, \ldots, \mathbf{x}_{N}$. The operator $\nabla_i^2$ is the Laplacian with respect to $\mathbf{x}_i$. The positive annealing function $\alpha(t)$ controls the relative weight of the kinetic and potential contributions. Equation~\eqref{eq:bare_QA} describes a real-time evolution; however, in practice, QA is often implemented in imaginary time rather than real time \cite{amara1993global,santoro2002theory,stella2005optimization,santoro2006optimization} (effectively replacing $i\partial_t$ with $-\partial_t$). Numerical studies indicate that this choice can lead to a quantitative improvement in performance, although the asymptotic scaling of the residual energy with annealing time remains unchanged~\cite{stella2005optimization}.

The annealing procedure is typically initialized in the ground state of a simpler Hamiltonian, after which $\alpha(t)$ is gradually decreased during the evolution.
The success of the algorithm relies on its ability to follow the instantaneous ground state, which is ensured for sufficiently slow annealing rates provided that a finite energy gap between the ground state and the excited states is maintained throughout the process \cite{born1928beweis}.
In the initial stages of QA, for large values of $\alpha$, the kinetic term dominates and the ground-state density is highly delocalized. In the opposite limit, $\alpha \to 0$, the Hamiltonian becomes dominated by the potential term, and the ground state approaches a localized function centered at the global minimum. From an intuitive perspective, QA can be interpreted as replacing Planck’s constant in the kinetic term with a time-dependent effective parameter $\tilde{\hbar}(t)=\sqrt{\alpha(t)}\,\hbar$, which controls the strength of quantum delocalization.

For many-body systems with continuous degrees of freedom, a direct implementation of Eq.~\ref{eq:bare_QA} is computationally demanding and typically requires the adoption of a simplifying ansatz for the wavefunction to keep the cost manageable~\cite{amara1993global}. Nevertheless, efficient approaches exist to sample the quantum nuclear density without explicitly representing the many-body wavefunction, as discussed in the following subsection.
\subsection*{Path-Integral Molecular Dynamics}
%\AFnote{Sezione un po' pedagogica. Si puo' accorciare a secondo della destinazione di pubblicazione.}
Among its many developments, Feynman’s path-integral formulation of quantum mechanics \cite{feynman1966quantum} provided a practical route to simulate nuclear quantum effects at finite temperature. Within this formalism, the quantum partition function of a system of distinguishable particles can be mapped onto an equivalent classical statistical mechanics problem:
%%%
\begin{align*}
    Z &= \int d\mathbf{x}^{(1)} \langle \mathbf{x}^{(1)} | e^{-\beta \hat H} | \mathbf{x}^{(1)} \rangle \\
      &= \lim_{P \to \infty} \int d\mathbf{x}^{(1)} \dots d\mathbf{x}^{(P)} 
      \prod_{j} \langle \mathbf{x}^{(j)} | e^{-\frac{\beta \hat H}{P}} | \mathbf{x}^{(j+1)} \rangle \\
      &= \lim_{P \to \infty} \int d\mathbf{x}^{(1)} \dots d\mathbf{x}^{(P)} 
      e^{-\frac{\beta}{P} U_P},
\end{align*}
%%%
where $\beta = (k_{\mathrm{B}} T)^{-1}$ is the inverse temperature, with $k_{\mathrm{B}}$ being the Boltzmann constant, and
%%%
\begin{align}\label{eq:U_P}
    U_P &= \sum_{i=1}^{N} \sum_{j=1}^{P} 
    \frac{1}{2} m_i \omega_P^2 \, 
    \big| \mathbf{x}_{i}^{(j)} - \mathbf{x}_{i}^{(j+1)} \big|^2
    + \sum_{j=1}^{P} U(\mathbf{x}_{1}^{(j)}, \ldots, \mathbf{x}_{N}^{(j)}),
\end{align}
\begin{align}\label{eq:omega_P}
    \omega_P = \frac{P k_{\mathrm{B}} T}{\hbar}.
\end{align}
%%%
Periodic boundary conditions, $\mathbf{x}_{i}^{(P+1)} = \mathbf{x}_{i}^{(1)}$, are imposed among the replicas of each particle $i$, commonly referred to as ``beads of the ring polymer'' in path-integral parlance.

In essence, the quantum problem is recast as the sampling of the canonical configurational space of a classical system composed of $N \times P$ particles interacting via the effective potential $U_P$. This effective canonical ensemble can be sampled using either Monte Carlo (MC) or molecular dynamics (MD) techniques, leading to the PIMC \cite{herman1982path,militzer2001path} and PIMD \cite{feynman1966quantum,ceriotti2014pi} approaches, respectively. Although both methods become exact in the limit of an infinite number of replicas, $P \to \infty$, in practice a finite number, typically $P \sim 10$--$100$, is sufficient for most applications.

In this work, we employ the PIMD approach in its ring-polymer molecular dynamics formulation, as implemented in the \textsc{i-PI} code \cite{ceriotti2014pi,kapil2019pi,litman2024pi}. The configurational space is sampled by numerically integrating the classical Hamiltonian
\begin{align*}
    H_P = \sum_{i=1}^{N} \sum_{j=1}^{P} 
    \frac{1}{2 m_i} \, \big| \mathbf{p}_{i}^{(j)} \big|^2 + U_P .
\end{align*}
Within this framework, quantum equilibrium averages can be obtained by averaging over the trajectory. The method is particularly straightforward for local observables $\hat{A}(x)$: 
\begin{align*}
    \langle \hat A(\mathbf{x})\rangle_Q &= \int d\mathbf{x}\langle \mathbf{x} | \frac{e^{-\beta \hat H}}{Z} \hat{A}| \mathbf{x} \rangle \\
    &= \frac{1}{P} \sum_{j=1}^{P} 
    \left\langle A(\mathbf{x}^{(j)}) \right\rangle ,
\end{align*}
where $\langle \cdot \rangle_{Q},\langle \cdot \rangle$ denote, respectively, the quantum average or the average of the corresponding classical observable over the PIMD trajectory. 

% For example, the quantum density can be expressed as
% \begin{align*}
%     \rho(\mathbf{r}) &= \langle \mathbf{r} | \frac{e^{-\beta \hat H}}{Z} | \mathbf{r} \rangle \\
%     &= \frac{1}{P} \sum_{j=1}^{P} 
%     \left\langle \prod_i \delta\!\left(\mathbf{x}_{i}^{(j)} - \mathbf{r}_i \right) \right\rangle ,
% \end{align*}
% where $\langle \cdot \rangle$ denotes an average over the PIMD trajectory.

Unlike (PI)MC methods \cite{herman1982path}, PIMD generates trajectories in a fictitious physical time. However, PIMD dynamics should be regarded as a tool to efficiently sample the effective canonical ensemble rather than as a means to reproduce exact quantum dynamics. Although approximate real-time dynamical information can, in some cases, be extracted from PIMD simulations, there are well-documented situations in which such results deviate from the true quantum dynamics \cite{perez2009comparative}. The dynamical aspects of QA-PIMD and their comparison with bare QA will be the subject of future investigation.

While the present paper focuses on the distinguishable particles formulation, the PIMD framework—and consequently the corresponding implementation of QA—can be generalized to efficiently incorporate bosonic exchange effects ~\cite{hirshberg2019path,feldman2023quadratic}.

\subsection*{Quantum, classical, and other flavors of annealing}

\begin{figure*}[!t]
    \centering
    \includegraphics[width=\linewidth]{ 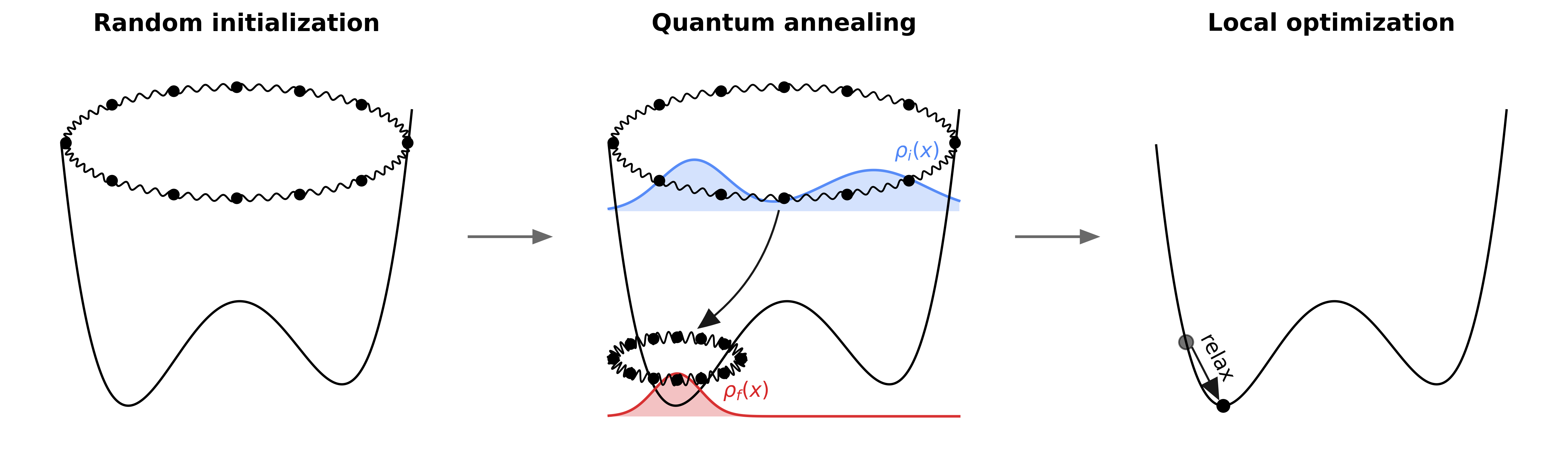}
    \caption{
    Schematic representation of the global optimization workflow. The process consists of three stages: (i) initialization, where the atomic positions of the replicas are randomly generated; (ii) quantum annealing, during which the ring-polymer density evolves from an initially delocalized distribution $\rho_i$ toward a localized final distribution $\rho_f$ as the control parameter is reduced; and (iii) local optimization, where residual thermal and/or quantum fluctuations are removed, yielding a well-relaxed structure.
    }
    \label{fig:workflow}
\end{figure*}

The implementation of quantum annealing through path-integral molecular dynamics is summarized in Fig.~\ref{fig:workflow} and resembles its classical counterpart, simulated annealing. Despite their different physical driving mechanisms, both approaches follow a similar optimization workflow, which can be naturally divided into three stages:
\begin{itemize}
    \item \textit{Initialization and equilibration}, during which random configurations are generated and rapidly equilibrated at large values of the control parameter to promote exploration.
    \item \textit{Annealing}, in which the control parameter is gradually reduced, steering the system toward low-energy configurations.
    \item \textit{Local optimization}, where the final structure obtained at the end of the annealing is refined to remove residual thermal and/or quantum fluctuations.
\end{itemize}

In SA, the system is initialized at high temperature, whereas in QA-PIMD it is prepared at large values of the effective quantum parameter $\tilde{\hbar}$. During annealing, the system is progressively ``cooled'' by decreasing the corresponding control parameter: the temperature $T(t)\to 0$ in SA and $\tilde \hbar(t)\to 0$ in QA-PIMD. As $\tilde{\hbar}$ decreases, the harmonic springs coupling the replicas stiffen according to
\begin{align}\label{omega_P_t}
    \tilde{\omega}_P(t) = \frac{P k_{\mathrm{B}} T}{\tilde{\hbar}(t)},
\end{align}
thereby reducing the spatial spread of the ring polymer around its centroid. In the absence of an interatomic potential, this quantum delocalization scales as~\cite{feynman1966quantum}
\begin{equation}\label{eq:deltax_debroglie}
    \Delta x \sim \frac{\hbar}{\sqrt{M k_{\mathrm{B}} T}}.
\end{equation}
While the above relation is not exact in the presence of a potential, imposing a $\Delta x$ comparable to the typical interatomic distance can still provide a rough estimate for the initial $\tilde{\hbar}$. After annealing, a brief local optimization step removes residual fluctuations and accelerates convergence. While this fine-tuning step is not conceptually required—since convergence could, in principle, be achieved through annealing alone—it is motivated by previous analyses showing the limited efficiency of both SA and QA in reaching the ``bottom" of harmonic wells ~\cite{santoro2002theory,santoro2006optimization}.

In addition to QA-PIMD, we also implement the replica-pinned variant of quantum annealing (RPQA), following Ref.~\cite{gregor2005minimization}. This strategy, previously introduced in the context of QA-PIMC, consists of periodically identifying and pinning the replica whose inherent structure (i.e., the structure obtained after a local optimization) exhibits the lowest potential energy. In practice, the QA simulation is temporarily paused at a prescribed pinning rate (e.g., every few picoseconds), and a local optimization is performed independently on all replicas. Then, as exemplified in Fig.~\ref{fig:summary_LJ} (d,e,f), the replica with the lowest-energy inherent structure is fixed in its relaxed configuration while the annealing continues. If a different replica subsequently attains a lower-energy configuration, the previously pinned replica is released and the new one is pinned. The process continues until the end of the annealing, when all replicas converge to the same minimum.

A key ingredient in all annealing strategies is the choice of the time schedule for the control parameters. For simplicity, unless specified otherwise, we adopt a linear schedule for both $T(t)$ in SA and $\tilde{\hbar}(t)$ in (RP)QA. The development of optimized or adaptive scheduling strategies is beyond the scope of the present work and will be addressed in future studies.

% \section{Numerical experiments}\label{sec:numerical_res} 
\subsection*{Asymmetric double-well potential}

Firstly, we benchmark the ability of QA-PIMD in reproducing and following the quantum nuclear density. We study one of the simplest, yet very educative, global optimization problems: the asymmetric double well (aDW)~\cite{finnila1994quantum,stella2005optimization,stella2006monte}. The one-dimensional potential is defined as
\begin{align*}
    U_{\mathrm{aDW}}(x) = -A(x-\delta)^2 + Bx^4,
\end{align*}
where $A$ and $B$ are positive parameters taken from Ref.~\cite{craig2005chemical}, and $\delta=0.2 \mathrm{~a.u.}$ introduces a slight asymmetry. The mass of the particles is set equal to hydrogen one.

Despite its simplicity, the aDW exemplifies many of the challenges encountered by global optimization algorithms. For instance, if the barrier height is sufficiently larger than the splitting between the two minima, pure QA [Eq.~\ref{eq:bare_QA}] is known to outperform SA by efficiently tunneling across the barrier~\cite{stella2005optimization}. In the present example, the temperature is a few times smaller than the splitting, which is roughly $7$ times smaller than the barrier.

\begin{figure}[!tb]
    \centering
    \includegraphics[width=\linewidth]{ 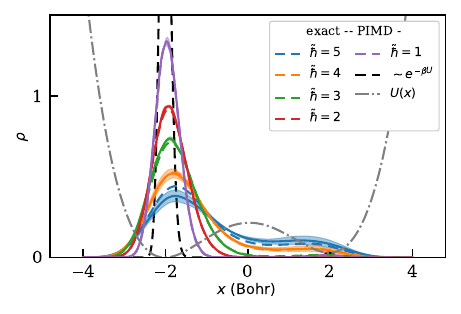}
    \caption{
    Local quantum density for an asymmetric double well potential for different values of $\tilde\hbar$, including the classical (formally $\tilde\hbar=0$) equilibrium density $e^{-\beta H}$. The dashed lines are the adiabatic densities at $T=100\mathrm{K}$ obtained from exact diagonalization. The continuous lines are obtained by $4$ $200\mathrm{ps}$-long QA-PIMD simulations with a linear annealing schedule for $\alpha(t)$. For each value of $\hbar$ and simulation, the density is obtained by averaging over a $4$ ps window around the corresponding $t$ such that $\tilde\hbar(\alpha( t))=\hbar$. The broadening is the standard deviation over 4 simulations.  The potential is shown in arbitrary units as a gray dashed line. 
    }
    \label{fig:aDW_QA}
\end{figure}

For the aDW system, we test QA-PIMD's ability to reproduce the adiabatic quantum density while being initialized with random positions for the replicas. In Fig.~\ref{fig:aDW_QA}, the local density obtained from PIMD simulations at different stages of the QA is compared to the exact adiabatic density, computed by direct diagonalization of \schro equation on a dense grid. The temperature dependence is straightforwardly included through the spectral representation 
\[
\rho(x)\propto \sum_n |\phi_n(x)|^2 e^{-\beta E_n},
\]
where $\phi_n(x)$ and $E_n$ are the eigenfunctions and eigenvalues of the Hamiltonian. The temperature is chosen to be smaller than the splitting between left and right minima and more than an order of magnitude smaller than the barrier.

The simulated quantum density overlaps remarkably well with the exact adiabatic result, even with relatively short trajectories and a reasonable number of beads ($P=64$). As expected, the quantum density evolves from being delocalized over the two wells to being localized in the deeper (left) one. Moreover, Fig.~\ref{fig:aDW_QA} shows a striking difference between the quantum and classical density near the barrier: at the given temperature, that region is practically ``classically prohibited'',
in the sense that it is thermally inaccessible on the timescale of the
corresponding classical simulation, while it is readily explored by the quantum
simulations with sufficiently large $\tilde{\hbar}$.

\subsection*{Optimization of Lennard--Jones clusters}

\begin{figure*}[!th]
	\centering
	\includegraphics[width=\linewidth]{ 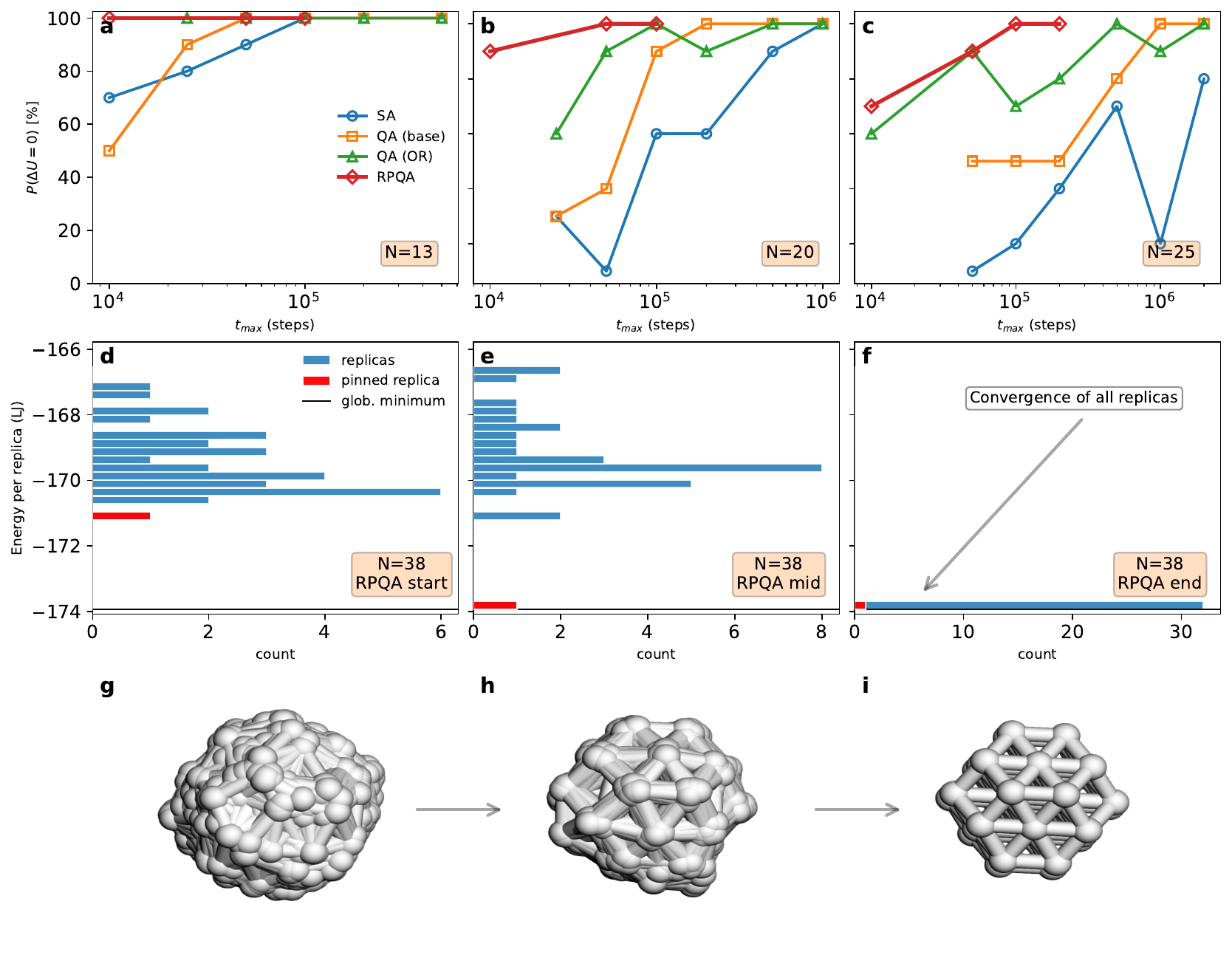}
    \caption{
    Global optimization of Lennard-Jones clusters with SA, QA-PIMD, and RPQA.
    (a--c) Probability of finding the reference global minimum~\cite{CCD} for $\mathrm{LJ}_{N}$ clusters with $N=13$, $20$, and $25$, respectively, using simulated annealing (SA), QA with base initialization, QA with optimal-ring initialization (OR), and replica-pinned quantum annealing (RPQA). Each point is computed from approximately $10$ independent annealing runs. Note the logarithmic scale of the x-axis.
    (d--f) Distribution of inherent-structure energies among the replicas during an RPQA simulation of $\mathrm{LJ}_{38}$, shown at $0$, $3\times 10^{4}$, and $2\times 10^{5}$ PIMD steps, respectively. The pinned replica is highlighted in red, while the horizontal black line indicates the global-minimum energy.
    (g--i) Representative stages of a RPQA-PIMD simulation, showing the progressive localization of the replicas during annealing.
    }
	\label{fig:summary_LJ}
\end{figure*}

This section focuses on the $\mathrm{LJ}_{N}$ problem, namely the determination of the optimal cluster configuration for $N$ atoms interacting via the Lennard--Jones (LJ) potential,
\begin{align*}
    U_{\mathrm{LJ}} = 4\epsilon_{\mathrm{LJ}}\sum_{j<i}\left[\left(\frac{\sigma}{r_{ij}}\right)^{12}-\left(\frac{\sigma}{r_{ij}}\right)^{6}\right],
\end{align*}
where $r_{ij}$ denotes the distance between atoms $i$ and $j$, while $\epsilon_{\mathrm{LJ}}$ and $\sigma$ define the characteristic energy and length scales, respectively. For generality, all results are reported in reduced LJ units; for example, the reduced temperature is defined as $T^* = k_{\mathrm{B}} T / \epsilon_{\mathrm{LJ}}$, and the reduced time as $t^* = t \sqrt{\epsilon_{\mathrm{LJ}}/(m \sigma^2)}$, where the atomic mass $m$ is set to the hydrogen mass.

Lennard-Jones clusters~\cite{hoare1976statistical,northby1987structure} represent a widely used benchmark for global optimization methods, including annealing-based techniques~\cite{amara1993global,finnila1994quantum,gregor2005minimization}, genetic algorithms~\cite{barron1999genetic,romero1999optimal}, and basin- and minima-hopping approaches~\cite{wales1997global,goedecker2004minima,krummenacher2024performing}. As the number of local minima grows rapidly with system size—ranging from an estimated $\sim 10^{3}$ for $\mathrm{LJ}_{13}$ to several hundred thousand for clusters with $N \sim 20$~\cite{hoare1976statistical,kostrowicki1991performance}—these systems constitute a stringent test for unbiased optimization algorithms. Reference energies are taken from the Lennard-Jones cluster database~\cite{CCD}, which compiles the best available estimates of the global minima. Although more stable structures could in principle be identified in future studies, we will refer to these reference energies as global minima throughout this work.

We first compare the performance of the different methods for three cluster sizes, as summarized in Fig.~\ref{fig:summary_LJ}(a--c). The panels show the probability of locating the global minimum ($\Delta U = 0$) as a function of annealing time. Each data point is obtained from statistics collected over several independent runs. For QA, two initialization protocols are considered: the base approach, where all replicas are initialized from the same atomic configuration, and the optimal-ring (OR) variant, where replicas are initialized from different inherent structures obtained from a liquid simulation. All QA simulations are performed at low temperatures, $T^* \approx 0.01$--$0.04$ (reduced LJ units), which are significantly smaller than the melting or evaporation temperatures typically used to initialize SA, $T^* \approx 0.3$--$0.35$. Additional technical details are provided in the Methods section and Supplemental Information (SI).

Figure~\ref{fig:summary_LJ}(a--c) shows a significant performance advantage of QA over SA: for a fixed annealing time, QA exhibits a higher probability of finding the global minimum than SA. This improvement is particularly pronounced for the OR initialization, which can be rationalized by its shorter equilibration time. Because the initial ring-polymer density is already delocalized over multiple basins, the OR protocol equilibrates more rapidly on average than the base initialization, in which all replicas initially lie within the basin of attraction of a single minimum. Unless stated otherwise, the OR initialization is employed in all path-integral simulations. While Fig.~\ref{fig:summary_LJ}(a--c) focuses on a few specific sizes, we also performed a broader analysis. In general, for both annealing strategies, larger clusters require longer annealing times to achieve the same probability of success as smaller systems, reflecting the exponential increase in the number of local minima. However, a few notable exceptions have been observed. For instance, the icosahedral cluster with $N=55$ is particularly easy to optimize, whereas $\mathrm{LJ}_{38}$ proves exceptionally challenging for both SA and QA and remains unsolved even with simulations extending up to $10^{6}$ steps. Overall, QA converges more rapidly than SA for clusters with $N \lesssim 30$, with a few outliers such as $N=27$, whose behavior is discussed later in the text.

The replica-pinned variant of quantum annealing leads to a further substantial enhancement of optimization performance, as shown in Fig.~\ref{fig:summary_LJ}. The replica-pinned strategy, originally introduced in the context of QA-PIMC~\cite{gregor2005minimization} and inspired by other evolutionary algorithms such as particle-swarm optimization~\cite{kennedy1995particle,bonyadi2017particle}, enables systematic tracking and refinement of the best structure encountered along the path-integral trajectory. Notably, RPQA efficiently solves the well-known $\mathrm{LJ}_{38}$ cluster, which appears beyond the reach of standard QA and SA methods and is regarded as a major challenge for unbiased global optimization algorithms~\cite{wales1997global}. The RPQA mechanism is illustrated in Fig.~\ref{fig:summary_LJ}(d--f). The inherent-energy distributions show that, after a few thousand PIMD steps, one replica reaches the basin of attraction of the global minimum and becomes pinned. While this replica remains fixed, the remaining replicas continue to explore higher-energy regions of configurational space and progressively converge toward the optimal structure as the annealing proceeds. This balance between exploration and exploitation is essential for systems characterized by double-funnel potential energy surfaces, such as $\mathrm{LJ}_{38}$~\cite{doye1999evolution}. Moreover, RPQA proves very effective even for the aforementioned outliers, e.g., $N=27$, as shown in the SI.

The replica-pinned approach entails an additional computational cost relative to standard QA, due to the extra local optimizations performed at each pinning event. Although the precise overhead depends on the details of the local optimization algorithm and the pinning rate, in most of our simulations the two components of the algorithm have comparable computational cost.

This overhead becomes negligible when shorter convergence times and the demonstrated ability to solve more complex clusters are taken into account (Fig.~\ref{fig:summary_LJ}). Moreover, despite the substantial methodological differences between Monte Carlo and molecular dynamics approaches, a qualitative performance comparison can be made with the basin-hopping algorithm~\cite{wales1997global}. The latter reports an $\sim 80\%$ success probability for $\mathrm{LJ}_{38}$ using approximately $5000$ local minimizations. In contrast, we obtain a comparable success probability using $32$ replicas and only $40$ local optimization rounds, performed once every $10^{4}$ steps.

Furthermore, because the local optimizations in RPQA are trivially parallelized across replicas, the effective computational speed-up seems substantial.

\subsection*{Crystal reconstruction}
\begin{figure*}[!tb]
    \centering
    \includegraphics[width=\linewidth]{ 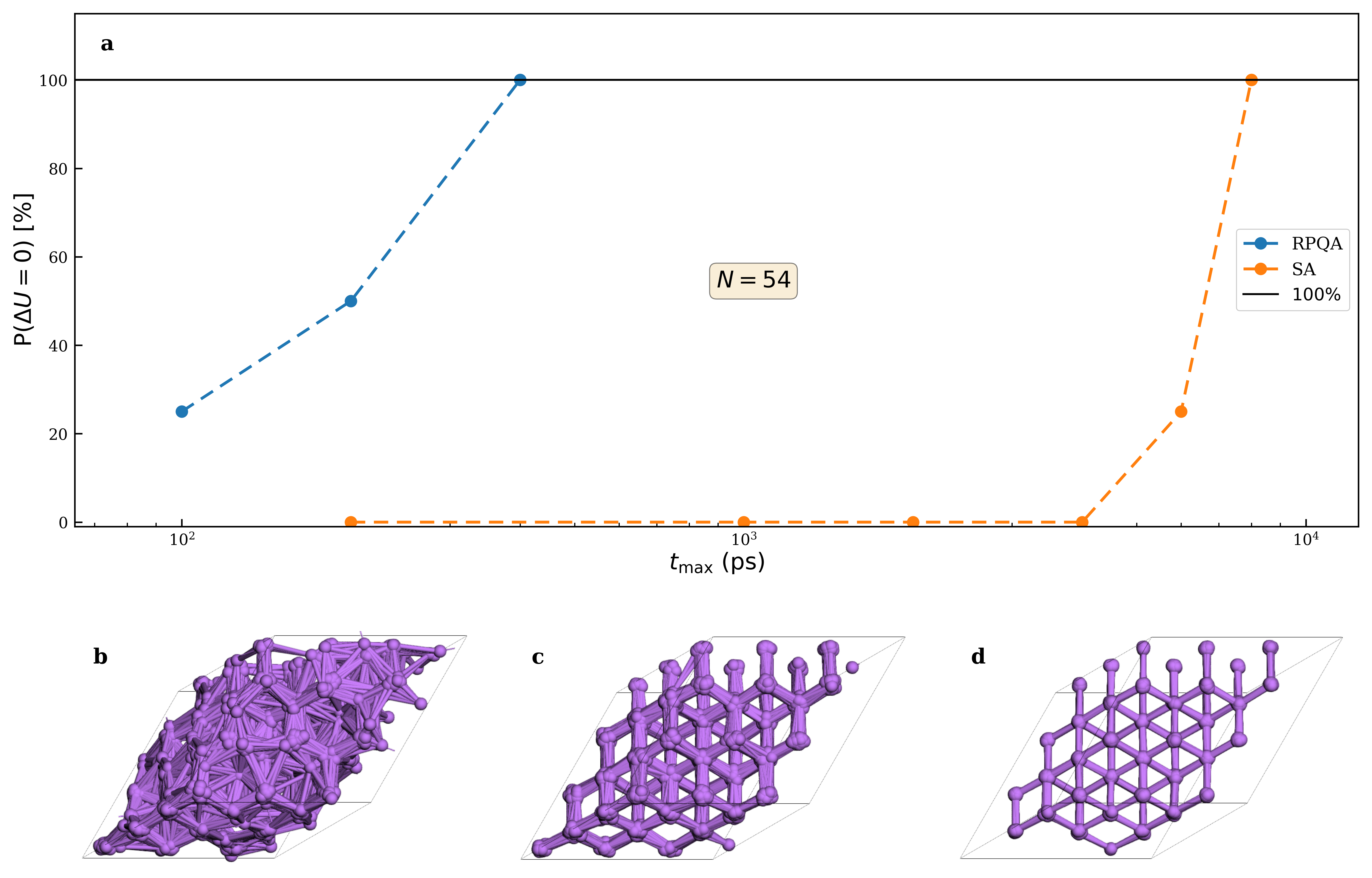}
    \caption{
    Reconstruction of a $54$-atom bulk silicon structure.
    (a) Probability of recovering the crystalline structure as a function of annealing time for both RPQA and SA, estimated from four independent runs for each annealing time. Note the logarithmic scale of the x-axis.
    (b--d) Snapshots of the RPQA simulation at different stages, progressively approaching convergence. $P=32$ replicas.
    }
    \label{fig:silicon}
\end{figure*}

Global optimization algorithms can also be used to reconstruct bulk materials when positional information is partial or missing. In the following, we consider two representative cases: the reconstruction of crystalline silicon and the determination of hydrogen positions in hydrides with known compositions.

The former provides a stringent test, as glass-forming materials such as silicon require very slow simulated annealing to avoid trapping in amorphous local minima. The latter case—missing hydrogen sites—illustrates a common experimental challenge: hydrogen atoms are notoriously difficult to detect using X-ray scattering techniques, often requiring neutron diffraction experiments~\cite{machida2014site} or refinement methods~\cite{woinska2016hydrogen}. As a result, databases such as MC3D contain a large number of structures with known chemical composition but uncertain hydrogen positions~\cite{huber2025mc3d,reents2026score}.

For silicon, the initial configurations are generated by randomizing the atomic positions of a crystalline structure in a $[3,3,3]$ supercell ($N=54$ atoms). The crystal structure is then reconstructed using both RPQA and SA. As shown in Fig.~\ref{fig:silicon}, RPQA consistently recovers the crystalline ground state even for annealing times more than an order of magnitude shorter than those required by the classical method. All simulations are performed using an existing MLIP~\cite{fan2022gpumd}, trained on crystalline, amorphous, and liquid silicon configurations.

For the latter problem, we focused on $10$ materials from the MC3D database and first verified that the reconstruction of missing hydrogen sites is genuinely a multi-minima global optimization problem. Indeed, a straightforward random search (RS), consisting of random hydrogen initialization followed by local optimization, sometimes fails to recover the lowest-energy structure even after $100$ independent attempts. In contrast, RPQA successfully reconstructs all systems using trajectories between $3$ and $40~\mathrm{ps}$, with a pinning rate of $1~\mathrm{ps}^{-1}$.

These calculations also show that the lowest-energy structure obtained with the MLIP does not always coincide with the database entry. For instance, for the $\mathrm{H}_6\mathrm{V}_{12}$ compound, RPQA identifies a structure with lower energy than the database structure even with the shortest trajectory, accompanied by a symmetry change from the original $\mathrm{Pmma}$ to $\mathrm{Pmm2}$. These results show that RPQA can efficiently solve experimentally relevant reconstruction problems without exploiting the extensive prior information available in the MC3D database~\cite{huber2025mc3d}. When inexpensive data-driven guesses are available, they could be included directly as one of the replicas in the initialization: the guess would remain pinned until, and unless, a lower-energy structure is found by the other replicas. Nevertheless, unbiased algorithms remain particularly valuable when prior information is limited, for example when exploring novel chemical compositions or extreme thermodynamic conditions.
\subsection*{(Quantum) free-energy: a thermodynamic perspective on the annealings }
\begin{figure*}[t!h]
    \centering
    \includegraphics[width=\linewidth]{ 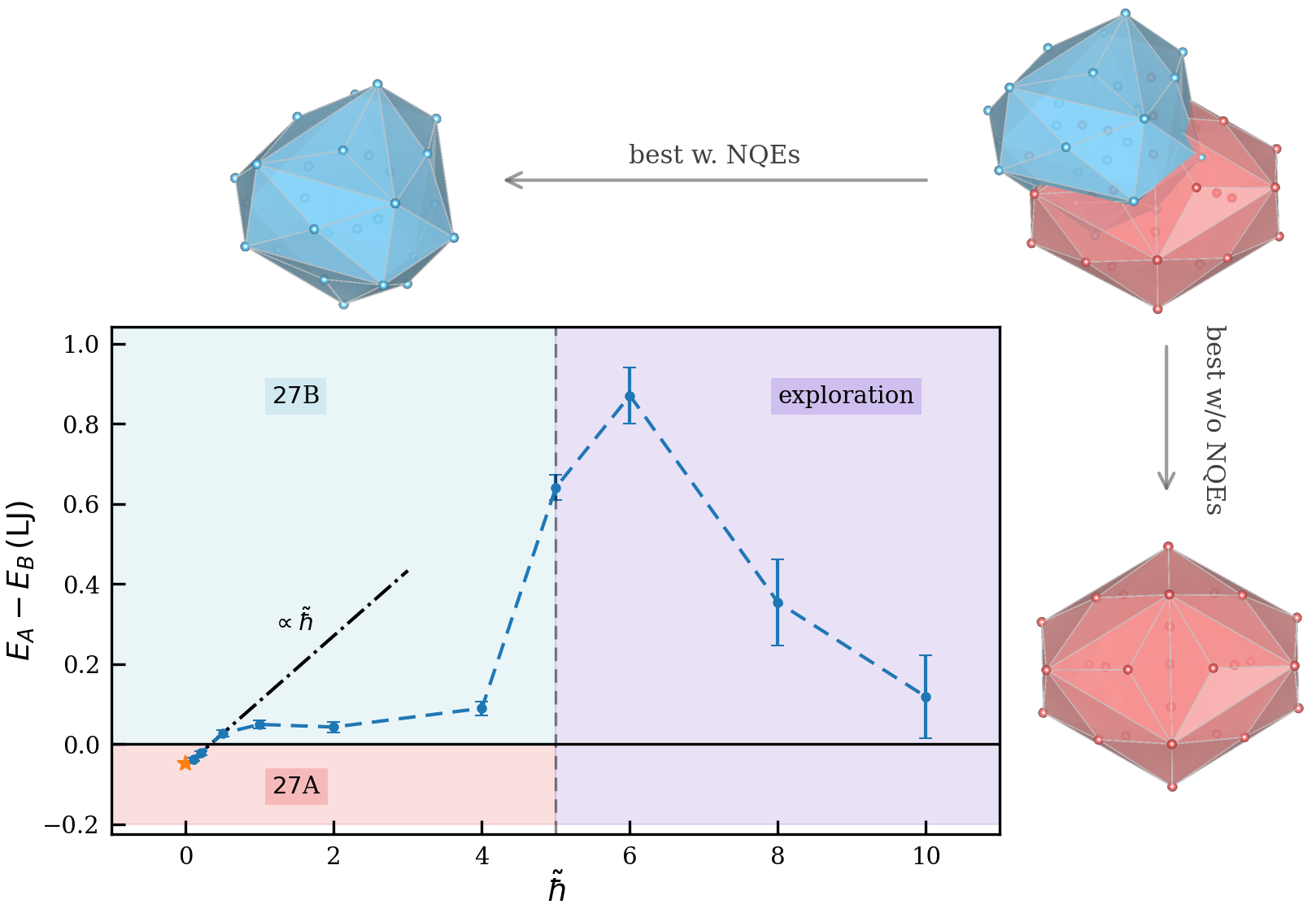}
    \caption{
    Quantum structural transition in $\mathrm{LJ}_{27}$ as a function of nuclear quantum effects.
    The central plot shows the energy difference between two competing LJ structures~\cite{calvo2001quantum}, computed from PIMD simulations at different values of $\tilde{\hbar}$. Positive values indicate stabilization of structure B relative to structure A.
    The surrounding structural diagrams show the competing minima and representative configurations extracted from simulations in the exploration regime.
    The energy is computed using the centroid estimator for the kinetic energy plus the average potential energy along the trajectory. All simulations were performed at $T^* \approx 0.01$, except for the star marker, which represents only the potential-energy difference between the two relaxed classical structures~\cite{calvo2001quantum}.
    }
    \label{fig:quantum_LJ_diagram}
\end{figure*}

The limitations, but also the potential applications, of both SA and QA-PIMD can be understood within a free-energy framework, following Ref.~\cite{finnila1994quantum}. Assuming ergodicity and sufficient equilibration time, (PI)MD samples a density that minimizes the (quantum) free energy,
\begin{align*}
    F &= E - TS, \\
    \rho_{\mathrm{eq}} &= \arg\min_{\rho} F[\rho].
\end{align*}
In both annealing schemes, the objective is to adiabatically follow the equilibrium density while evolving a control parameter, either $\tilde{\hbar}$ or $T$. However, this objective may be obstructed by ergodicity breaking or long equilibration times, potentially leading to suboptimal results.

For simulated annealing, temperature-driven structural transitions can severely hinder performance. Consider a critical temperature $T_c$ such that, for $T > T_c$, the structure minimizing the free energy differs from the global minimum (GM) of the potential energy surface (PES), while for $T < T_c$ the diffusivity $D$ becomes extremely small. In such a scenario, SA may fail to follow the phase transition and remain trapped in a metastable basin. A representative example is $\mathrm{LJ}_{38}$, where temperature-induced structural transitions complicate optimization~\cite{calvo2001quantum}.

To investigate the corresponding quantum case, we consider the $\mathrm{LJ}_{27}$ cluster. According to Ref.~\cite{calvo2001quantum}, this system undergoes a quantum structural transition: when nuclear quantum effects such as zero-point energy (ZPE) are included, the most stable structure changes once the system becomes sufficiently quantum. For example, lighter atoms such as Neon favor a different structure than heavier, chemically similar atoms such as Argon. We refer to the structure minimizing the quantum free energy as the \emph{quantum minimum}. Although we focus on a single cluster size, Ref.~\cite{calvo2001quantum} reports similar quantum transitions in roughly one third of rare-gas clusters with $N < 100$.

To characterize this transition, we compute the energy difference between the two competing structures, A (classical minimum) and B (quantum minimum), as a function of $\tilde{\hbar}$ (Fig.~\ref{fig:quantum_LJ_diagram}). For each value of $\tilde{\hbar}$, PIMD simulations are performed at very low temperature, $k_{\mathrm{B}}T/\epsilon_{\mathrm{LJ}} \approx 0.01$, such that entropic contributions are assumed to be negligible. In the classical limit, structure A has lower energy. However, as the quantumness increases, structure B becomes more energetically favorable, with a transition occurring around $\tilde{\hbar} \approx 0.3$.

For small $\tilde{\hbar}$, the energy difference varies linearly, consistent with the harmonic approximation for the ZPE:
\begin{align*}
    \Delta \mathrm{ZPE}
    &= \mathrm{ZPE}_{A} - \mathrm{ZPE}_{B} \\
    &= \frac{\tilde{\hbar}}{2} \sum_{\mu} \left( \omega_\mu^{A} - \omega_\mu^{B} \right),
\end{align*}
where $\omega_\mu^{A,B}$ denote the vibrational frequencies of the two structures.

For larger $\tilde{\hbar}$, the dependence becomes nonlinear as PIMD captures anharmonic quantum effects beyond the harmonic approximation, while structure B remains energetically preferred. At sufficiently large $\tilde{\hbar}$, replicas occupy different minima at the end of the simulations; we refer to this regime as the \emph{exploration} region. In this regime, comparisons between individual structures become less meaningful, as the sampled density spans multiple basins of the PES.

On this system, QA and RPQA exhibit markedly different behavior. In the exploration regime, the classical minimum is frequently visited, allowing RPQA to pin it efficiently. In contrast, QA typically converges to the quantum minimum, even when the final $\tilde{\hbar}$ lies below the transition value. A quantitative analysis is reported in the SI. This behavior mirrors the difficulty encountered by SA when a structural transition occurs at very low temperature, where thermal fluctuations become ineffective at exploring the PES.

While the existence of quantum minima may hinder QA when the goal is to minimize the classical PES, it also represents an opportunity. Many materials—particularly those containing light atoms and/or under high pressure—exhibit phases stabilized by nuclear quantum effects. In conventional approaches, quantum minima are often obtained \emph{a posteriori} by adding ZPE corrections to classically optimized structures. In contrast, QA naturally incorporates NQEs during the search, allowing the physically relevant structure to be recovered simply by terminating the annealing when $\tilde{\hbar}(t) = 1$.
\subsection*{Quantum structural transition of high-pressure $\mathrm{LaH}_{10}$}

\begin{figure*}[th!] \centering \includegraphics[width=1\linewidth]{ 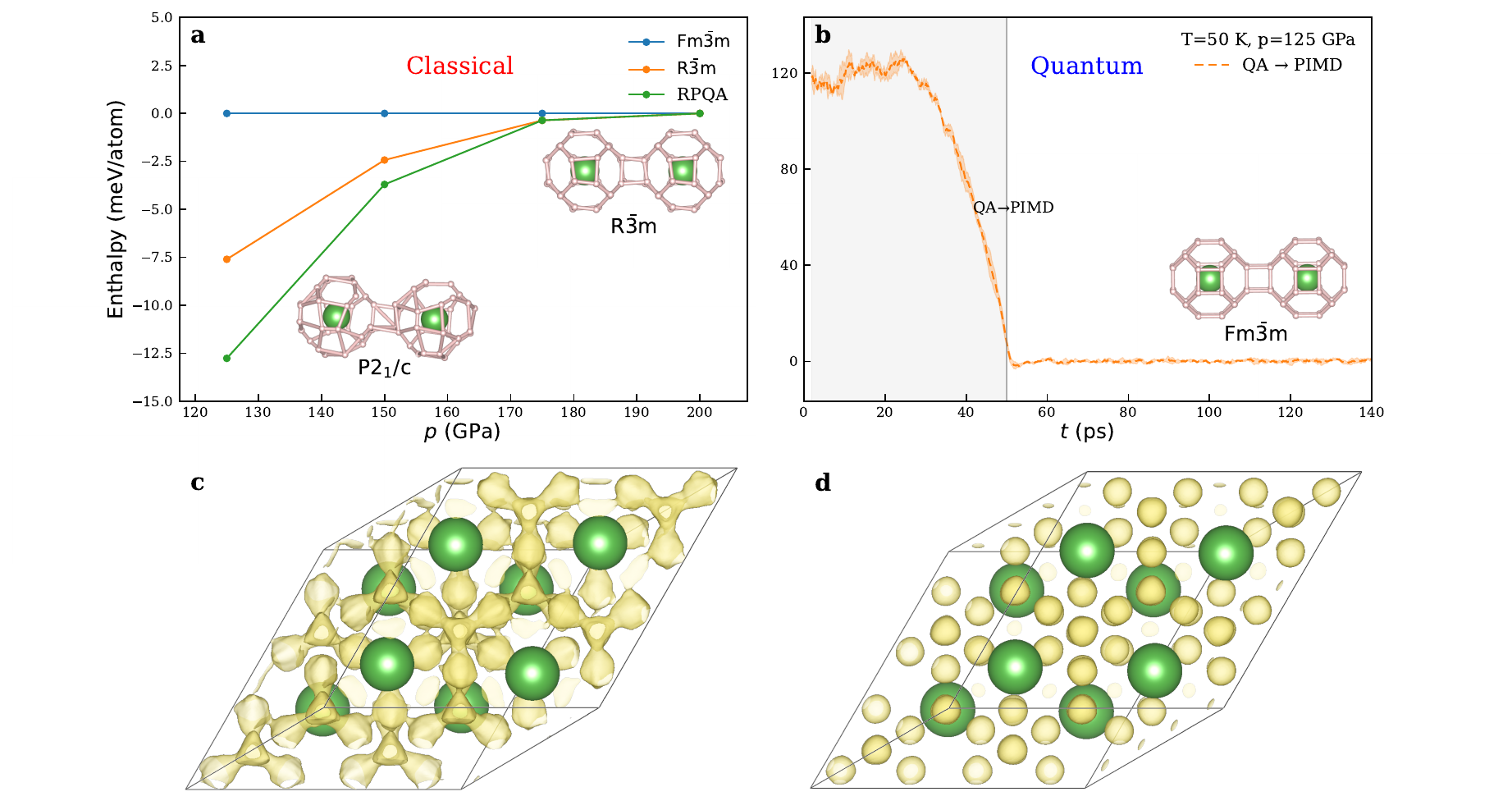} 
    \caption{
    Stability of $\mathrm{LaH}_{10}$ at high pressures.
    (a) Classical enthalpy curves at $T=0~\mathrm{K}$ for different crystal structures, compared with RPQA results.
    (b) Quantum enthalpy obtained from QA runs of $50~\mathrm{ps}$ followed by PIMD equilibration at $\tilde{\hbar}=1$. In both panels, the enthalpy of the $Fm\bar{3}m$ phase, classical or quantum respectively, is taken as reference.
    (c,d) Isosurfaces of the hydrogen density at the initial and final stages of the QA simulation, corresponding to $\tilde{\hbar}=6$ and $\tilde{\hbar}=1$, respectively.
    All QA and RPQA simulations were performed at $T=50~\mathrm{K}$ using $P=32$ replicas.
    }
\label{fig:quantum_LaH10_diagram} 
\end{figure*}

The phase diagram of the high-pressure superconductor $\mathrm{LaH}_{10}$ provides a striking example of the importance of including nuclear quantum effects. According to Refs.~\cite{errea2020quantum,poletaev2025sscha,smith2026capturing}, incorporating NQEs substantially modifies the phase diagram of the high-symmetry $\mathrm{Fm\bar{3}m}$ structure, leading to much closer agreement with experimental observations~\cite{drozdov2019superconductivity}. Classically, this phase is not the global enthalpy minimum below $p \approx 200~\mathrm{GPa}$ and is even mechanically unstable at $p=125~\mathrm{GPa}$ according to frozen-phonon calculations~\cite{errea2020quantum}. However, when NQEs are treated within the Stochastic Self-Consistent Harmonic Approximation (SSCHA)~\cite{monacelli2021stochastic}, the $\mathrm{Fm\bar{3}m}$ phase becomes the most stable among the structures considered, even at $p=125~\mathrm{GPa}$~\cite{errea2020quantum}.

To investigate this system efficiently, we trained a MLIP. Starting from the MACE foundational model~\cite{batatia2025foundation}, its accuracy was significantly improved through additional fine-tuning on Density Functional Theory (DFT) data. Model validation and a phonon analysis confirming the classical instability of the $\mathrm{Fm\bar{3}m}$ phase are reported in the SI.

We then explored the classical and quantum phase diagrams using RPQA and QA, respectively. The results at $p=125~\mathrm{GPa}$ are shown in Fig.~\ref{fig:quantum_LaH10_diagram}. Panel (a) reports the classical enthalpy curves at $T=0~\mathrm{K}$, obtained from structural relaxations of known phases and from RPQA. For pressures $\lesssim 200~\mathrm{GPa}$, RPQA correctly identifies lower-symmetry structures whose classical enthalpy lies below that of the $\mathrm{Fm\bar{3}m}$ phase.

In contrast, the quantum simulations (panel b) confirm the stability of the high-symmetry phase. The quantum enthalpy was evaluated during QA runs in which the control parameter was gradually reduced to its physical value, $\tilde{\hbar}(t)=1$, followed by standard PIMD simulations. The time-averaged structure along the PIMD trajectories corresponds to the $\mathrm{Fm\bar{3}m}$ one, demonstrating both its quantum stabilization and the ability of QA to locate the quantum minimum even when initialized from different symmetries and/or with additional disorder. 

During the initial stage of the QA, the hydrogen density is strongly delocalized (Fig.~\ref{fig:quantum_LaH10_diagram}c), reflecting efficient exploration of the configuration space. As $\tilde{\hbar}$ approaches its physical value, the density progressively localizes (panel d), yielding the stabilized high-symmetry structure. Similar behavior is observed for several pressures in the range $125 < p < 200~\mathrm{GPa}$ and for temperatures $T=50$ and $100~\mathrm{K}$.

In summary, while RPQA effectively minimizes the classical potential (enthalpy) surface, the resulting structure does not necessarily correspond to the physically relevant phase. The $\mathrm{Fm\bar{3}m}$ structure is stabilized by NQEs and can be naturally recovered by ``stopping'' the QA procedure when quantum fluctuations reach their physical value. In contrast to approaches that first identify candidate structures through classical optimization and subsequently incorporate NQEs via phononic free-energy calculations, QA directly explores the quantum free-energy landscape. Therefore, QA-PIMD may provide an effective strategy for investigating high-pressure hydrides such as the present one or $\mathrm{H}_3\mathrm{S}$~\cite{errea2016quantum}, where nuclear quantum effects are known to induce a quantum structural transition.

\section{Conclusions}\label{sec:conclusions}
In conclusion, this work presents and critically analyzes a novel implementation of (RP)QA using PIMD. Thanks to the use of PIMD, this global optimization method maintains the simplicity and computational cost of a few molecular dynamics simulations, avoiding the huge hurdle of working on the quantum wavefunction or making simplifying assumptions on its functional form~\cite{amara1993global}. Moreover, many of the tools of PIMD, like constraints, barostats or metadynamics~\cite{quhe2015path}, have been or can be integrated into this technique, increasing the flexibility of the method and its ability to explore specific regions of the phase diagram.

As a global optimization method, (RP)QA-PIMD shows remarkable performance among unbiased structure-search approaches. On the standard benchmark of Lennard-Jones clusters, QA-PIMD often reaches the target structure with shorter annealing schedules than SA. In particular, the replica-pinned variant~\cite{gregor2005minimization} consistently solves the challenging $\mathrm{LJ}_{38}$ cluster with relatively short simulations, which is a notable result for an unbiased method. The origin of this speed-up is likely two-fold. First, quantum fluctuations allow the system to access regions of configuration space that are thermally inaccessible, or only very rarely sampled, by the corresponding classical annealing protocol, as illustrated by the numerical experiments on the asymmetric double well in Fig.~\ref{fig:aDW_QA}. Second, the path-integral representation naturally introduces a set of communicating replicas, enabling strategies such as replica pinning, which is reminiscent of population-based optimization methods such as particle swarm algorithms~\cite{bonyadi2017particle}. For most of the systems considered here, PIMD simulations of only a few tens of thousands of steps are sufficient to reliably identify the best structure found in our searches. This timescale is practically affordable when combined with MLIPs and with the efficient parallelization over replicas implemented in modern PIMD frameworks~\cite{litman2024pi}. As exemplified by
the $\mathrm{LaH}_{10}$ case, fine-tuned MLIPs can reproduce phase-stability trends in excellent agreement with \emph{ab initio} calculations~\cite{errea2020quantum}.

Finally, we emphasize the double-edged role of quantum structural
transitions in quantum annealing. On the one hand, the presence of such
transitions may help rationalize difficulties encountered in global
optimization using quantum-annealing schemes without replica
pinning~\cite{amara1993global,finnila1994quantum,gregor2005minimization}.
On the other hand, the ability of QA-PIMD to explore the quantum
free-energy landscape provides a powerful route to structure prediction in
materials with strong NQEs~\cite{calvo2001quantum,errea2020quantum,
monacelli2023quantum,poletaev2025sscha,smith2026capturing}.

% \subsubsection*{SI Datasets}

% \paragraph*{SI Movies}

\subsection*{Computational details}\label{app:computational}
 All (PI)MD simulations are performed using a modified version of the i-Pi code~\cite{ceriotti2014pi}. Depending on the material, we used internal i-Pi drivers, e.g. aDW or LJ, or the ASE~\cite{ase-paper} interface for MACE or NEP machine-learning potentials~\cite{batatia2025foundation,fan2022gpumd}. The timestep for MD has been chosen to guarantee a sufficient convergence of energy in an NVE simulation, $dt=0.005t^*$ for LJ systems, $dt=2~\mathrm{fs}$ for silicon, and $dt=1~\mathrm{fs}$ for all the other systems. Before any annealing, either classical or quantum, an equilibration run of $10^4$ timesteps is performed. To guarantee the correct integration of high-frequency modes among replicas, i-Pi multistep functionality has been used: the (almost inexpensive) harmonic forces between beads are computed with a time-step $10-20$ smaller than the classical forces.

The main cost of a $P$-bead PIMD step is the evaluation of the
interatomic potential and its forces for each bead, amounting to roughly $P$ independent force evaluations. These evaluations are independent across beads and are therefore naturally parallelizable. In the i-PI implementation, the ring-polymer replicas can be distributed over independent force-evaluation clients, so the relevant practical cost is the wall-clock overhead at fixed computational allocation, together with the resource-normalized cost when additional nodes are used. In the present $P=16,32$ calculations, the typical fixed-resource wall-clock overhead relative to classical MD is substantially smaller than the nominal factor $P$, and bead-level parallelism can further reduce the wall-clock penalty. Further details on the initialization of the PIMD simulations are provided in the SI.

 Ab-initio DFT calculations are performed using  the Quantum ESPRESSO code~\cite{giannozzi2009quantum,giannozzi2017advanced,giannozzi2020quantum} through the workflow manager AiiDA~\cite{huber2020aiida,uhrin2021workflows}.
 
 Finally, chemiscope~\cite{Fraux2020} and VESTA~\cite{momma2008vesta} have been used for visualization. 
 
 Further technical details and scripts are provided below.

\subsection*{Beads and temperature convergence}\label{app:P_T_convergence}
In the following section, we study the role of temperature and the number of beads $P$ in a QA-PIMD simulation.

The temperature acts in the PIMD simulations in a two-fold way. Firstly, the quantum density at finite temperature contains excited states, weighted by the Boltzmann factor. As a rule of thumb, the temperature should be lower than the average splitting between the low-energy minima. Secondly, the temperature alters the interaction between beads, through Eq.~\ref{eq:omega_P}, yielding classical simulations at sufficiently high temperatures (all replicas are strongly coupled, no quantum delocalization).

Regarding the number of beads, it should be increased with the degree of quantumness, with observables typically converging with some power-law behavior $O(1/P^n)$~\cite{ceriotti2014pi}. 
%However, the quantumness is not constant in a QA simulation, as it goes from ``super-quantum" ($\tilde \hbar>1$) to classical $\tilde \hbar\to 0$. It is not yet studied how an uncertainty in the initial phase of the annealing propagates in the final results. For instance, in Fig.~\ref{fig:aDW_QA}, the $\tilde \hbar=5$ local density slightly differs from the exact one, but the subsequent ones perfectly overlap the exact adiabatic one. Intuitively, as long as the initial density has a sufficient overlap with the ground state and the annealing is sufficiently slow, the algorithm should be able to relax to the correct density. 

\begin{figure}[t]
	\centering
	\includegraphics[width=\linewidth]{ 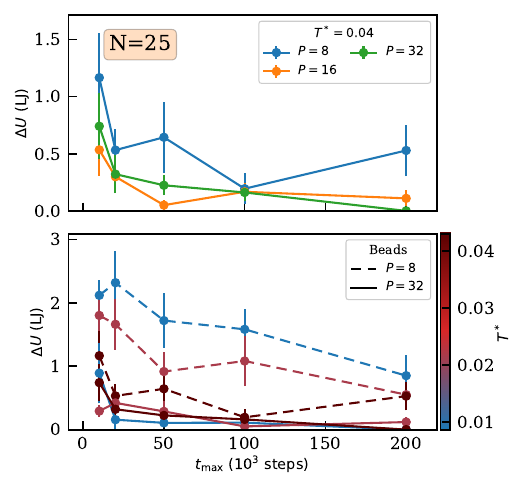}
	\caption{  QA (OR) performance for $\mathrm{LJ}_{25}$ as a function of the annealing time, for different numbers of beads $P$ and temperatures. Top panel: fixed temperature, $T^*=0.04$ K, and  $P=8,16,32$. Bottom panel: $P=8$ (dashed line) and $P=32$ for different temperatures. }
	\label{fig:LJ_PT_convergence}
\end{figure}

In order to find a suitable set of parameters, the QA's performance is studied for different numbers of replicas and different temperatures, as exemplified in Fig.~\ref{fig:LJ_PT_convergence} for a LJ cluster. The simulations with just $P=8$ have an overall inferior performance to the $P=16,32$ ones, especially at lower temperatures. Overall, a typical number of beads for PIMD simulations, $P=32$, seems to be already at convergence, even for low temperatures.

% \subsection*{Subsection for Method}
% Example text for subsection.

\subsection*{Data availability}\label{sec:code_and_data}
The codes that support the relevant results within this paper are publicly available from the respective developers' repositories. 
The (RP)QA-PIMD algorithm is implemented in a modified version of the I-Pi code\cite{ceriotti2014pi,kapil2019pi,litman2024pi}, publicly accessible on GitHub upon publication. Templates and pre-/post-processing Python scripts are available on the Materials Cloud platform~\cite{talirz2020materials}. See DOI:[to be included when available].

\subsection{Acknowledgement} The authors are grateful to S. Baroni, L. Monacelli, M. Ceriotti, D. Tisi, A. Carta, L. Voj\'{a}\v{c}ek, P. Settembri, M. Gubler, and F. Grasselli for insightful discussions. This research was supported
by the NCCR MARVEL, a National Centre of Competence in Research, funded by the Swiss National Science Foundation (grant number 205602). Computer time was provided
by the Swiss National Supercomputing Centre (CSCS)
under project No. lp18, and mr33.
%V. Sanella, S. Sch\"{a}ren

% \bibsplit[0]
%Use \bibsplit to split the references from the body of the text. Value "[3]" represents the number of reference in the left column (Note: Please avoid single column figures & tables on this page.)

% Bibliography
\bibliography{main}

\end{document}